\documentclass[a4paper, 12pt]{article}

\usepackage{graphicx}
\usepackage{url}
\usepackage{amsbsy,amssymb} 
\usepackage{amsmath}
\usepackage{abstract}
\usepackage{hyperref}
\usepackage{color}

\usepackage{amsmath,amsthm}

\usepackage[all]{xy} 

\newtheorem{Proposition}{Proposition}
\newtheorem{Corollary}{Corollary}

\newtheorem{Definition}{Definition}
\newtheorem{Example}{Example}

\begin{document}

{\LARGE\centering{\bf{Inverse problem in the calculus of variations - functional and antiexact forms}}}

\begin{center}
\sf{Rados\l aw Antoni Kycia$^{1,2,a}$}
\end{center}

\medskip
 \small{
\centerline{$^{1}$Masaryk University}
\centerline{Department of Mathematics and Statistics}
\centerline{Kotl\'{a}\v{r}sk\'{a} 267/2, 611 37 Brno, The Czech Republic}
\centerline{\\}
\centerline{$^{2}$Cracow University of Technology}
\centerline{Department of Computer Science and Telecommunications}
\centerline{Warszawska 24, Krak\'ow, 31-155, Poland}
\centerline{\\}

\centerline{$^{a}${\tt
kycia.radoslaw@gmail.com} }
}

\begin{abstract}
\noindent

We connect the well-known theory of functional forms of variational bicomplex with the theory of antiexact differential forms. We identify antiexact functional forms as an obstruction to the variationality of differential equations. The most prominent result of this observation is the formulation of the variational problem for some differential equations that are not variational and neither have a variational multiplier. Heat, Navier-Stokes, and KdV equations illustrate this new variational principle.

\end{abstract}
Keywords: functional forms; Euler-Lagrange equation; inverse problem in the calculus of variations; variationality; Poincar\'{e} lemma; antiexact forms; homotopy operator \\
Mathematical Subject Classification:  49-02, 49N99 \\

\section{Introduction}

The calculus of variations is a unique bridge between various mathematics disciplines such as differential geometry and functional analysis. It also has an unprecedented role in physics, being a universal language for Field Theory and various problems in other areas of science. Due to its importance, it was developed into a vast discipline, and there have been developed many approaches to this subject, including the traditional way in terms of functional analysis \cite{GelfandFomin, VariationsHilderbendt, Vainberg}, and the purely geometrical approach in terms of jet spaces and variational bicomplex introduced in \cite{variationalbicomplexVinogradov, variationalbicomplexTulczyjew, variationalbicomplexTsujishita}, see also the pedagogical introduction in \cite{AndersonVariations, Krupka, Krupkova, HandbookOfGlobalAnalysisKrupka, HandbookOfGlobalAnalysisVitolo, HandbookOfGlobalAnalysisKrupkkova, SaundersSummary, JetSaunders, Olver, Khavkine}; for applications in physics, see a review in, e.g., \cite{Khavkine2, Olver, AndersonVariations, Krupkova}.

One of the fundamental directions in the calculus of variations is to know if given differential equations can be reformulated as a variational problem. This is the main subject of so-called inverse problems (IP) in the calculus of variations \cite{ZenkovInverseProblem, Olver, Markus}, see also a formulation in terms of Exterior Differential Systems \cite{AndersonInverse, PrinceInverse2, PrinceInverseproblem}. We assume we do not modify equations - we take them 'as they stand'. The set of Helmholtz conditions allows one to check the variationality. In the more relaxed version of this problem, one initially allows one to multiply the equations by some function or perform a linear transformation of the equations to make them variational if there are not variational as they stand \cite{SaundersSummary, ZenkovInverseProblem}. Currently the IPs are considered in various situations as Lie groups \cite{ZoltanLieGroupsIP} or in non-smooth setup in control theory, e.g., \cite{FedericoIPNonsmooth}. Moreover, some ideas on variationality, as discussed below, are used in controlability of PDEs \cite{Control1, Control2}.

IP has a long history initiated by Helmholtz \cite{Helmholtz} and carried out by many researchers, see \cite{SantilliMechanics} for the early history of IP. We will focus on geometric methods, the summary of which was provided in, e.g., \cite{SaundersSummary, ZenkovInverseProblem, Olver, HandbookOfGlobalAnalysisKrupka, HandbookOfGlobalAnalysisKrupkkova, HandbookOfGlobalAnalysisVitolo, SantilliMechanics}. The most advanced exploration in this direction is the variation bicomplex \cite{AndersonVariations, Olver, HandbookOfGlobalAnalysisKrupka, HandbookOfGlobalAnalysisVitolo, Krupka} which one flavor useful in applications is based on the notion of Functional Forms, e.g., \cite{Olver}, i.e., application of exterior calculus to functionals that appear in the calculus of variations. One of the crucial steps in solving 'locally' IP is the notion of adjointness of some differential operator on jet space. This notion in \cite{SantilliMechanics} was elevated to the fundamental principle of mechanics. The related concept (symmetry of Gateaux derivative on suitable Banach space) in functional-analytic approach to IP was developed in \cite{Vainberg} (Theorem 5.3). The anti-self-adjointness also has profound meaning in the Hamiltonian approach to IP \cite{Olver, Samoilenko}

The exterior calculus on Functional Forms also diffused to physics leading to formulation proposed by Aldrovandi and Kraenkel in \cite{FunctionalExteriorCalculus} and \cite{FunctionalExteriorCalculus-Book}, chapter 20. For some equations it sometimes leads to easier and more straightforward computations for local IP solutions than, yet equivalent to, the one originating from the Helmholtz conditions \cite{FunctionalExteriorCalculus-Book}. 

The above results are primarily global. The local elaboration in a star-shaped region uses a (vertical) homotopy operator that is analogous to the one from the proof of the Poincare lemma \cite{Olver, KyciaPoincare}. However, to our best knowledge, it was not noticed that the homotopy operator has properties that can be used to be explored in the calculus of variations further in this local setup. In particular, it defines a new class of forms - antiexact forms \cite{EdelenExteriorCalculus, EdelenIsovectorMethods, KyciaPoincare, KyciaPoincareCohomotopy} which analogs will be identified in the variational calculus in this paper. The paper presents a novel approach to the local theory of IP using these properties. It is organized as follows: In the next section, we discuss the known results and conventions used in the calculus of functional forms and define the properties of antiexact forms. Then we use these results to identify antiexact functional forms as an obstruction to IP. Thanks to this, it is possible to provide the most important result of this paper - to formulate a more general variational problem that includes problems for the equations that are not variational as they stand. Some applications of this new variational principle to essential equations such as heat, Navier-Stokes, and KdV equations are provided.

\section{Related results}
This section will summarize the results of the theory of functional forms and the local theory of homotopy operator related to antiexact forms.

\subsection{Functional forms}
\label{Subsection_FunctionalForms}
The theory of functional forms is presented following \cite{Olver} (section 5.4); see also \cite{FunctionalExteriorCalculus, FunctionalExteriorCalculus-Book, JetSaunders}. 

Since our considerations will be local, as usually in applications, we restrict ourselves to a subset of the product manifold $M\subset X \times U$, where $X$ will be called a horizontal (base) space and $U$ a vertical space. It can be interpreted as (a local trivialization of) a bundle $\pi: X \times U \rightarrow X$. Moreover, we assume that $U$ is star-shaped, i.e., $M$ is vertically star-shaped. By $J\pi$ denote the infinite jet \cite{JetSaunders} of smooth sections $X\rightarrow U$.

Let $\mathcal{G}$ be an algebra of smooth functions over $J\pi$. The coordinates for $X$ are $\{x^{i}\}_{i=1}^{n}$, for $U$ are $\{u^{\alpha}\}_{\alpha=1}^{k}$ and jet coordinates are $(x^{i},u^{\alpha},u^{\alpha}_{J})$ for a multiindex $J=(i_{1},\ldots,i_{p})$ with $i_{l}\in \{1,\ldots, n\}$. For a smooth function $f:X\rightarrow U$, its holonomic lift to infinite jet bundle is denoted by $jf$.

Let $\omega = \sum_{J} P_{J}[u]dx^{J}$ be a form with $P_{J} \in \mathcal{G}$. The total differential of $\omega$ is
\begin{equation}
 D\omega = \sum_{i=1}^{n} D_{i}P_{J} dx^{i}\wedge dx^{J},
\end{equation}
where $D_{i} = \frac{\partial}{\partial x^{i}} + \sum_{\alpha, J} u^{\alpha}_{J,i}\frac{\partial}{\partial u^{\alpha}_{J}}$ is a total derivative with respect to $x^{i}$.

Our main interest are vertical forms represented in local coordinates as
\begin{equation}
 \hat{\omega} = \sum_{J, \alpha} P^{J}_{\alpha}[u] du_{J_{1}}^{\alpha_{1}}\wedge\ldots \wedge du_{J_{p}}^{\alpha_{p}},
\end{equation}
where $\alpha=\alpha_{1}\ldots \alpha_{p}$. The vertical derivative $\hat{d}$ is defined locally as
\begin{equation}
 \hat{d}\hat{\omega} = \sum_{\alpha, J, \beta, K} \frac{\partial P_{\alpha}^{J}}{\partial u^{\beta}_{K}} du^{\beta}_{K}\wedge du_{J_{1}}^{\alpha_{1}}\wedge\ldots \wedge du_{J_{p}}^{\alpha_{p}}.
\end{equation}
The vertical $k$-forms form a vector space $\hat{\Lambda}^{k}$.

We have also useful commutativity relation $ D_{i}\hat{d}\hat{\omega} = \hat{d}D_{i}\hat{\omega}$.

A functional (0-form) is defined as functional over $\mathcal{G}$ integrated symbolically over a base space. Their vector space will be denoted by $S\mathcal{G}$. The numerical value of the function is obtained by pulling back the integrand using some holonomic section $\Phi$ (prolongation of functions $X\rightarrow U$ to the jet space $J\pi$), namely,
\begin{equation}
 S[\Phi] = \int_{\Omega} \Phi^{*}L[u] dx \in \mathbb{R},
\end{equation}
where $\Omega \subset X$ is some compact set, and $L\in \mathcal{G}$ depends on the jet coordinates collectively written as $u$.

It is useful to introduce equivalence relation between integrands by 
\begin{equation}
 \tilde{L} \sim L \quad \Leftrightarrow \quad \tilde{L} = L + Div P,
\end{equation}
where $Div ~ P = \sum_{i} D_{i}P$ for some $P\in \mathcal{G}$. Then the functionals defined by $L$ and $\bar{L}$ are the same assuming, hereafter, that $\Phi^{*}P[u] = 0$ on $\partial \Omega$. Therefore we can define a space of integrands $\mathcal{F}=\mathcal{G} / \sim $. Then the functional (0-form) is a symbolic integral over a representative of an element of $\mathcal{F}$. They will be denoted as elements of $S\mathcal{F}$.

Likewise, one can define a similar equivalence on the space of vertical forms $\hat{\Lambda}^{k}$, by 
\begin{equation}
 \hat{\omega}=\hat{\omega}'+Div ~ \hat{\eta}, \quad \hat{\eta} \in \hat{\Lambda}^{k},
\end{equation}
that provides
\begin{equation}
 \Lambda_{*}^{k} := \hat{\Lambda}^{k} / Div(\hat{\Lambda}^{k}).
\end{equation}
Then the functional $k$-form is the integral over the base space of the representative of this form. The space of functional $k$-forms will be denoted by $S\Lambda_{*}^{k}$.

The alternating structure anticommutes with total derivative be means of integration by parts and using vanishing of boundary term, i.e.,
\begin{equation}
 \int \hat{\omega}\wedge (D_{i} \hat{\eta}) dx = -\int (D_{i}\hat{\omega})\wedge \hat{\eta}dx, \quad \hat{\omega}\in \hat{\Lambda}^{k}, \quad \hat{\eta}\in \hat{\Lambda}^{l}.
\end{equation}


In physical literature to avoid expansion of notation, so called, 'continuum Einstein convention' (see \cite{FunctionalExteriorCalculus-Book}, section 20.2.8) or (super)condensed notation notation \cite{deWitt} is used, that means we omit the integral sign when writing functional differential form, i.e.,
\begin{equation}
 \int (u_{xx} du) dx \rightarrow u_{xx}du.
\end{equation}
This convention annihilates differences in writing elements from $\Lambda_{*}$ and $S\Lambda_{*}$. However, in this paper, to avoid ambiguity in formulas, we will keep the integral over the base manifold using 'operator notation' (measure before the integrand):
\begin{equation}
 \int dx f(x,u).
\end{equation}

Functional forms form only a vector space over $\mathbb{R}$ and not the module over $C^{\infty}$-functions on $X$. Therefore, when we multiply a functional form by a function or the other form, we mean it at the level of integrands.

\subsection{Inverse problem of the calculus of variations}

If we have Euler-Lagrange equations $\{E_{\alpha}[u]=0\}_{\alpha}$, then we can construct an Euler form $\Omega=E_{\alpha}du^{\alpha}$.

We introduce, following \cite{Olver} Section 5.4, the canonical form for the first order functional form
\begin{equation}
 \Omega = \int dx E_{\alpha}[u]du^{\alpha},
 \label{Eq_EulerForm}
\end{equation}
as well as the canonical form for the second order functional form
\begin{equation}
 F = \int dx \sum_{\alpha, \beta}\mathcal{D}_{\alpha \beta}(du^{\alpha})\wedge du^{\beta},
\end{equation}
where $\mathcal{D}_{\alpha\beta}$ is a differential operator.

Any vertical vector field $X$ is called a variation and can be inserted into an Euler functional form to provide a numerical value for a functional.

From the fundamental theorem of the (local) inverse problem of the calculus of variations, it results that the set of the equations $\{E_{\alpha}\}_{\alpha}$ is a set of Euler-Lagrange equations iff the vertical derivative of their Euler form vanish \cite{Olver} (or equivalently when the Frechet derivative of the Euler form vanishes, see \cite{Vainberg}, Theorem 5.3). This means that $\mathcal{D}_{\alpha\beta}$ is self-adjoin. This condition can be cast into the classical Helmholtz conditions. It is essential to underline that the equations are used as they stand, i.e., without any modifications.

There is also a modified inverse problem of the calculus of variations -  the multiplier problem: whether there exists a multiplier to the Euler form making this $1$-functional form exact \cite{ZenkovInverseProblem, SaundersSummary}. The solutions can be reduced to solving some system of differential equations for the multiplier function that results from the Helmholtz conditions.

Moreover, the Euler form must be constructed using the proper ordering of equations in (\ref{Eq_EulerForm}), which gives rise to the ordering problem \cite{SaundersSummary}. We can introduce a matrix $A$ that rearrange the order of equations in the construction of the Euler form and that gives, if possible, an exact form $\Omega =\int dx \sum_{\alpha\beta} E_{\alpha}A^{\alpha}_{\beta}du^{\beta}$. Then the form of the matrix $A$ is provided by solving $\hat{d}\Omega=0$.

\subsection{Antiexact differential forms}
\label{Appendix_AntiexactForms}
In this section we summarize some notions from \cite{EdelenExteriorCalculus, EdelenIsovectorMethods, KyciaPoincare, KyciaPoincareCohomotopy, CopoincareHamiltonianSystem} for the reader convenience.

In what follows de Rham complex is defined on a star-shaped open subset $V$ of $\mathbb{R}^{n}$, which can be transformed diffeomorphically to an open subset of a manifold. Then the exterior derivative $d$ has a corresponding homotopy operator $H$, such that they fulfill the homotopy formula
\begin{equation}
 dH + Hd = I - s_{x_{0}}^{*},
 \label{Eq_homotopyFormula}
\end{equation}
where $s_{x_{0}}^{*}$ is the pull-back on the single-point manifold $x_{0}$, and $I$ is the identity. The homotopy  operator for $\omega \in \Lambda^{k}(V)$ is 
\begin{equation}
H\omega = \int_{0}^{1} \mathcal{K}\lrcorner\omega|_{F(t,x)}t^{k-1}dt, 
\end{equation}
where $\mathcal{K}=(x-x_{0})^{i}\partial_{i}$, and $F(t,x)=x_{0}+t(x-x_{0})$ is the linear homotopy between $x \in V$ and the center $x_{0}\in V$ of this homotopy. This specific homotopy operator for linear homotopy is nilpotent $H^{2}=0$, as it was noted in \cite{EdelenExteriorCalculus} and used recently in \cite{KyciaPoincare, KyciaPoincareCohomotopy, CopoincareHamiltonianSystem}.

For a star-shaped region $V$ every closed form is exact, or equivalently the cohomology group $H^{p}(V)=0$ for $p > 0$.

It occurs that the image of the homotopy operator induces a module of antiexact forms \cite{EdelenExteriorCalculus, EdelenIsovectorMethods, KyciaPoincare} defined as 
\begin{equation}
\mathcal{A}(V) = \{ \omega \in \Lambda(V) | \mathcal{K}\lrcorner \omega = 0, \omega|_{x=x_{0}}=0\}. 
\end{equation}
They, with a vector space of exact/closed forms (on a star-shaped $V$) $\mathcal{E}(V)=\{\omega \in \Lambda(V)|d\omega = 0\}$, introduce the direct sum decomposition
\begin{equation}
 \Lambda(V) = \mathcal{E} \oplus \mathcal{A}. 
 \label{Eq_ExactAntiexactDecomposition}
\end{equation}
This can be extended to codfifferential for Riemannian manifolds \cite{KyciaPoincareCohomotopy, CopoincareHamiltonianSystem}.

Then the projection operators utilizing (\ref{Eq_homotopyFormula}) are as follows
\begin{equation}
 Hd:\Lambda(V)\rightarrow \mathcal{A}(V), \quad Hd: \Lambda(V)\rightarrow \mathcal{E}(V).
\end{equation}
This also gives that $\mathcal{A}=Ker(H)$, and $\mathcal{E}=Ker(d)$ in a star-shaped region.

These concepts allow us to solve (locally) various equations of mathematical physics \cite{KyciaPoincareCohomotopy}.

One of the most fundamental equations in geometry governing (local) obstruction for integrability of many structures (e.g., integrability of horizontal distribution) is the following \cite{EdelenExteriorCalculus}
\begin{equation}
 d\Omega = \Gamma \wedge \Omega + \Sigma',
 \label{Eq_ObstructionEquation}
\end{equation}
for $\Gamma \in \Lambda^{1}(V)$, $\Omega \in \Lambda^{k}(V)$, and the torsion is $\Sigma \in \Lambda^{k+1}(V)$. Using the decomposition (\ref{Eq_ExactAntiexactDecomposition}) we can write $\Gamma=d\gamma + \theta$, where $\theta \in \mathcal{A}^{1}(V)$, which gives
\begin{equation}
 d\Omega = d\gamma\wedge \Omega + \Sigma,
\end{equation}
where $\Sigma = d\gamma + \theta$. Then the solution is \cite{EdelenExteriorCalculus}
\begin{equation}
 \begin{array}{ll}
  \Omega = e^{\gamma}(d\phi + \eta), & \Sigma = e^{\gamma}d\eta \\
  \phi = H(e^{-\gamma}\Omega), & \eta = H(e^{-\gamma}\Sigma)=Hd(e^{-\gamma}\Omega).
 \end{array}
\end{equation}
Therefore, when $\Sigma =0$, we have that $\Omega$ is up to multiplicative factor an exact form.

\section{Antiexact functional forms}
This section presents our results relating the local theory of antiexact forms and its relation to the calculus of variations. The setup will be a product bundle $\pi: X\times U \rightarrow X$ and its infinite jet bundle $J\pi$, as it was presented in subsection \ref{Subsection_FunctionalForms}.

Starting from an Euler functional $1$-form $\Omega = \int dx E_{\alpha}du^{\alpha}$, its vertical derivative $\hat{d}$ in the canonical form is
\begin{equation}
 \hat{d}\Omega = \int dx\frac{\partial E_{\alpha}}{\partial u^{\beta}_{K}}D_{K}(du^{\beta})\wedge du^{\alpha} = \int dx \mathcal{D}_{\alpha\beta}(du^{\alpha})\wedge du^{\beta}.
\end{equation}
Decomposing $\mathcal{D}$ operator into self-adjoint part $\mathcal{D}^{s}$ and anti-self adjoint part $\mathcal{D}^{a}$ we have that the antisymmetry of the integration by parts and wedge product results with
\begin{equation}
 \hat{d}\Omega = \int dx \mathcal{D}^{a}_{\alpha\beta}(du^{\alpha})\wedge du^{\beta}.
 \label{Eq_antiselfadjointOf2Form}
\end{equation}

The vague statement is that the functional forms approach to the IP can lead faster to the check of variationality ('as it stands') than the Helmholtz conditions. It is so because the cancelation of some terms in $\hat{d}\Omega$ may occur before extracting the anti-self adjoint part of $\mathcal{D}$ due to additional antisymmetry arising from the wedge product. The Helmholtz conditions are (anti)symmetrized version of the operator in $\hat{d}\Omega$ that results from integration by parts and linear combinations of terms. If the $\hat{d}\Omega$ vanishes at some stage of such operations, then it vanishes for all equivalent elements/representatives in $S\Lambda_{*}$.

By choosing some center of homotopy $u_{0}$ in the jet of sections $J\pi$ we make by standard construction (as in \cite{Olver}) the (vertical) homotopy operator $H$ for $\hat{d}$. For a functional form $A = \int dx P_{\alpha}^{J}du^{\alpha_{1}}_{J_{1}}\wedge\ldots\wedge du^{\alpha^{p}}_{J_{p}}$, the (vertical) homotopy $F(t,x,u) = u_{0} + t(u-u_{0})$, and $\mathcal{K}$, being the prolongation of $K=(u-u_{0})^{\alpha}\partial_{u^{\alpha}}$ to the infinite jet, we have,
\begin{equation}
 HA = \int dx \int_{0}^{1}\mathcal{K}\lrcorner (P_{\alpha}^{J}du^{\alpha_{1}}_{J_{1}}\wedge\ldots\wedge du^{\alpha^{p}}_{J_{p}})|_{F(t,x,u)} t^{p-1}dt.
\end{equation}

We can now define (compare with the definition \ref{Appendix_AntiexactForms} of ordinary differential forms)
\begin{Definition}
 Antiexact functional forms $S\hat{\mathcal{A}}$ are functionals of vertical forms in the image of the projection operator $H\hat{d}$.
\end{Definition}

Similarly as for the de Rham complex on a star-shaped domain, see \cite{EdelenExteriorCalculus, EdelenIsovectorMethods, KyciaPoincare}, we have a decomposition of vertical forms in a star-shaped $U$, given by 
\begin{equation}
 \hat{\Lambda}^{k}=\hat{\mathcal{E}}^{k}\oplus\hat{\mathcal{A}}^{k},
\end{equation}
where $\hat{\mathcal{E}} = \{\omega\in \hat{\Lambda} | \hat{d}\omega =0\}$ are vertical exact and hence closed forms. The projector to $\hat{\mathcal{E}}$ is $\hat{d}H$. This extends to the functionals of these forms
\begin{equation}
 S\hat{\Lambda}^{k}=S\hat{\mathcal{E}}^{k}\oplus S\hat{\mathcal{A}}^{k},
\end{equation}
by the linearity of the integral over the base space $X$. We have a similar decomposition for the quotient spaces by div-equivalence and for their functional spaces.

Then we have the following proposition 
\begin{Proposition}
 The antiexact functional $1$-forms are the homotopy image of (\ref{Eq_antiselfadjointOf2Form}).
\end{Proposition}
It means that in the case of $S\hat{\Lambda}^{1}$ (or $S\hat{\Lambda}_{*}^{1}$) the antiexact forms are homotopy image of (\ref{Eq_antiselfadjointOf2Form}).

One can use the obstruction equation (\ref{Eq_ObstructionEquation}) to formulate 
\begin{Proposition}
 If the exterior derivative of the Euler form can be rewritten as
 \begin{equation}
  \hat{d}\Omega = \hat{d}\gamma\wedge \Omega + \Sigma
 \end{equation}
with the vanishing torsion form $\Sigma=0$, then there is a variational multiplier $e^{\gamma}$ such that
\begin{equation}
 \Omega = e^{\gamma}\hat{d}\phi
\end{equation}
for some vertical $0$-from $\gamma$.
\end{Proposition}

\section{Failing of differential equations to be variational as a new optimization problem}

This section will provide a possible redefinition of the variational problem for differential equations that are not variational as they stand nor have multipliers.

As a motivating example we consider the well-known case of the heat equation $E[u]=u_{t}-u_{xx}=0$ defined on $J(\mathbb{R}^{2}\rightarrow \mathbb{R})$. This gives the rise to the Euler form $\Omega = \int dx (u_{t}-u_{xx})du$. It is known that one can make it variational by simply adding another equation $v_{t}+v_{xx}=0$ in a new variable $v$, which extends the space to $J(\mathbb{R}^{2}\rightarrow \mathbb{R}\times\mathbb{R})$. This trick can be seen as an attempt to remove the non-variational term $u_{t}du$ in the original equation. The enhanced system induces the Euler form $\bar{\Omega}=\int dx [(u_{t}-u_{xx})dv+(v_{t}+v_{xx})du]$, which now fulfills $\hat{d}\bar{\Omega}=0$. It occurs that this adjoint equation can be interpreted as a control to the original heat equation \cite{Control2, Control1}. That rise the question if there is another way to represent lack of variationality for heat and other important equations.

We will consider an approach that does not involve modifying the jet space of solutions, however, it imposes a constraint on possible variations.

Starting from the equations $\{E_{\alpha}[u]=0\}$, we construct the Euler form $\Omega=\int dx E_{\alpha}[u]du^{\alpha}$. We can decompose the Euler form into $S\hat{\mathcal{E}}^{1}\oplus S\hat{\mathcal{A}}^{1}$ as
\begin{equation}
 \Omega = \alpha + \beta, \quad \alpha \in S\hat{\mathcal{E}}^{1},~ \beta \in S\hat{\mathcal{A}}^{1}.
\end{equation}
The element $\alpha$ can be written locally as a vertical derivative of some functional 
\begin{equation}
 \alpha = \hat{d}\int dx L[u].
\end{equation}
The element $\beta$ can be written locally as a homotopy image of some element from $S\hat{\mathcal{A}}^{2}$ in the form 
\begin{equation}
 \beta = H\int dx \mathcal{D}^{a}_{\alpha\beta}(du^{\alpha})\wedge du^{\beta}.
\end{equation}
We have then the following optimization problem
\begin{Definition}
\label{Def_optimizationProblem}
 Write the optimization problem as a direct sum of elements from $S\hat{\mathcal{A}}^{0} \oplus S\hat{\mathcal{E}}^{2} \subset S\hat{\Lambda}^{0} \oplus S\hat{\Lambda}^{2}$, where an element from $S\hat{\mathcal{A}}^{0}$ is a functional 0-form, and an element from $\hat{\mathcal{E}}^{2}$ is related by a homotopy operator to the obstruction part from $\hat{\mathcal{A}}^{1}$. The problem has the form  
\begin{equation}
 S[u] = \int dx L[u]  \oplus \int dx \mathcal{D}^{a}_{\alpha\beta}(du^{\alpha})\wedge du^{\beta}.
\end{equation}
Then the modified variation is represented by the $\delta_{2} = \hat{d} \oplus H$ and gives
\begin{equation}
 \delta_{2} S[u] = \underbrace{\hat{d} \int dx L[u] }_{\alpha} \oplus \underbrace{H\int dx \mathcal{D}^{a}_{\alpha\beta}(du^{\alpha})\wedge du^{\beta}}_{\beta}.
\end{equation}
 Then the optimization problem is to find a solution $u$ and a variation $X\neq 0$ such that
 \begin{equation}
  (ju)^{*}(X \lrcorner \beta) =0, \quad \forall u
 \label{Eq_FirstProblem}
 \end{equation}
and 
\begin{equation}
 (ju)^{*}(X \lrcorner \alpha) =0,
 \label{Eq_SecondProblem}
\end{equation}
for all $X$ fulfilling (\ref{Eq_FirstProblem}).

The structure of the optimization problem is pictured in the diagram in Fig. \ref{Fig.DecompositionFull}.
 \begin{figure}
\centering
\xymatrix{ 0 & \ar[l]_{\hat{d}} \ldots \ar@/^/[drr]^{H}  & \oplus & \ldots \ar[r]^H & 0 \\
0  & \ar[l]_{\hat{d}} S\hat{\mathcal{E}}^{2}   \ar@/^/[drr]^{H}        &  \oplus  & \ldots \ar[r]^H \ar@/^/[ull]^{\hat{d}} & 0  \\
0 & \ar[l]_{\hat{d}} S\hat{\mathcal{E}}^{1} \ar@/^/[drr]^{H}  & \oplus & S\hat{\mathcal{A}}^{1} \ar[r]^H \ar@/^/[ull]^{\hat{d}}  & 0 \\
0 & \ar[l]_{\hat{d}} \mathbb{R} & \oplus & S\hat{\mathcal{A}}^{0} \ar@/^/[ull]^{\hat{d}}  \ar[dl]^H   &  \\
     &  &   0   &
}
\caption{Relations between different parts of the optimization problem. The bottom level represents the functional as an element of $S\hat{\mathcal{A}}^{0}$ defined up to constant, the Euler form as an element of $S\hat{\mathcal{E}}^{1}$, the image of the homotopy map of the obstruction as an element of $S\hat{\mathcal{A}}^{1}$, and an obstruction to the variationality as an element of $S\hat{\mathcal{E}}^{2}$. It is a relevant part of the full diagram presented in \cite{KyciaPoincare}, Fig. 1.}
\label{Fig.DecompositionFull}
\end{figure}
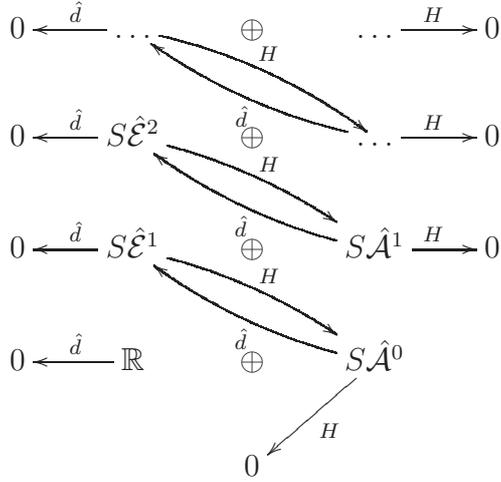
\end{Definition}

From (\ref{Eq_FirstProblem}) we get all the possible variations $X$, and from (\ref{Eq_SecondProblem}) we get, for these specific variations the solutions for $u$. In this sense both equations are dual.

The problem (\ref{Eq_SecondProblem}) distinguishes some specific $u$, under the assumption given by the following
\begin{Proposition}
 The optimization problem from the Definition \ref{Def_optimizationProblem} is well-defined when $X \notin Ker (\alpha)$. 
\end{Proposition}

One can look at the optimization problem $\Omega=\alpha+\beta$ from the Definition \ref{Def_optimizationProblem} the other way round. We restrict the possible variations $X$ to such a subspace that $X \lrcorner (\alpha+\beta) = X\lrcorner \alpha$. This leads to the following conclusions
\begin{Corollary}
 Starting from a functional $S$, we can extend optimization problem given by this functional $\hat{d}S = \alpha =0$ by adding an antiexact functional form $\beta \in S\hat{\mathcal{A}}^{1}$ with the constraint for variation $X$ given by $X\lrcorner \beta=0$, as long as, $X \notin Ker(\alpha)$.
\end{Corollary}
This extension generates a whole class of problems that differ from each other by an antiexact functional $1$-form.

We illustrate this issue with well-definiteness in a few examples.
\begin{Example}
 First we start with the heat equation $E[u]=u_{t}-u_{xx}=0$ in $1+1$ dimension. We reformulate the problem as an optimization problem for 
\begin{equation}
 S=\int dx \frac{1}{2} u_{x}^{2},
\label{Eq_HeatFunctional1}
\end{equation}
with the additional constraint that for the variational field $X=jY$ being the jet prolongation of $Y=\eta \partial_{u}$ fulfills
\begin{equation}
 X\lrcorner (u du_{t}) =0,
\end{equation} 
that is,
\begin{equation}
 \eta_{t}=0.
 \label{Eq_ConstraintOnVariationHeat}
\end{equation}
That means that variational vector fields are prolongations of $Y=\eta(x)\partial_{u}$.

We have $\alpha = \int dx u_{xx}du$, $\beta=-\int dx u_{t}du$.

This constraint on variation (\ref{Eq_ConstraintOnVariationHeat}) does not affect optimization of the functional (\ref{Eq_HeatFunctional1}) due to different variable in differentiation.
\end{Example}

Therefore, in this case, we can define the following problem
\begin{Example}
 In the case of the heat equation we have
\begin{equation}
 S[u] =  \int dx \frac{1}{2}(u_{x})^{2} \oplus \int dx \frac{1}{2}(du_{t}\wedge du),
\end{equation}
which upon $\delta_{2}$ provides the variation
\begin{equation}
 \delta_{2}S [u] = \int dx (u_{t}-u_{xx})du,
\end{equation}
where the irrelevant constant term was omitted.
\end{Example}

\begin{Example}
 As a negative example we use the slightly modified equation
\begin{equation}
 u_{x} - u_{xx}=0.
\end{equation}
Then the obstruction term in the Euler form is $u_{x} du$, which induces variations $X \neq 0$ that fulfills $\eta_{x}=0$. The functional is again (\ref{Eq_HeatFunctional1}), as before. However, now $\hat{d}S = \int dx u_{xx}du = -\int dx u_{x}du_{x}$, that vanish on these specific variations, and therefore, allows all possible solutions for $u$. This can be also easily seen if we rewrite the Euler form as $E[u]=-\int dx (u-u_{x})\partial_{x}du$.
\end{Example}

\begin{Example}
Another example are the Navier-Stokes equations (see, e.g., \cite{FunctionalExteriorCalculus-Book}, section 20.3.3) are defined for sections of $\mathbb{R}^{4}\times J\mathbb{R}^{3+1}$ where the coordinates on the base manifolds are $t,\{x\}_{i=1}^{x}$ and field variables are $\{v_{i}\}_{i=1}^{3}$ (velocity field) and $p$ (pressure). They are of the form
 \begin{equation}
  \rho (\partial_{t}v_{i}+ v_{j}\partial^{j}v_{i})+\partial_{i}p-\mu \partial^{j}\partial_{j}v_{i}=0,
  \label{Eq_NSEquation}
 \end{equation}
with incompressibility condition $\partial_{i}v^{i}=0$. The velocity fields form a vector in a Euclidean space.

The Euler form, taking into account that $p$ is a Lagrange multiplier, is
\begin{equation}
 \Omega = \int dx\left[ \rho \partial_{t}v_{i}dv^{i}+\rho (v_{j}\partial^{j}v_{i})dv^{i}+ \hat{d}\left(\frac{1}{2}\mu(\partial_{j}v_{i}\partial^{j}v^{i}-p(\partial_{j}v^{j}) \right)  \right].
\end{equation}
Then the first two terms do not vanish under vertical derivative, and so they are an obstruction to variationality.

The variational vector field $X=jY=j(\eta^{i} \partial_{v^{i}}+\chi\partial_{p})$ we obtain the following constraints
\begin{equation}
 \partial_{t}\eta^{i}=0, \quad v_{i}v_{j}\partial^{j}\eta^{i}=0,
\end{equation}
or equivalently
\begin{equation}
 \partial_{t}\eta^{i}=0, \quad v_{i}\partial^{j}(v_{j}\eta^{i})=0,
\end{equation}
for all $v^{i}$. That shows that $\eta=\eta(x^{i})$ and there is a nontrivial conditions for gradients of $\eta^{i}$. 

Then the generalized variational problem is as follows
\begin{equation}
 S=\int dx \left(\frac{1}{2}\mu(\partial_{j}v_{i}\partial^{j}v^{i}-p(\partial_{j}v^{j}) \right) \oplus \hat{d}\int dx\left( \rho \partial_{t}v_{i}dv^{i}+\rho (v_{j}\partial^{j}v_{i})dv^{i}\right).
\end{equation}
\end{Example}

\begin{Example}
 The final example is KdV equation \cite{FunctionalExteriorCalculus-Book} with the Euler form
 \begin{equation}
  \Omega = \int dx \left(u_{t}+\partial_{x}\frac{u^{2}}{2}+u_{xxx}\right)du.
 \end{equation}
Since the equation contains only even derivatives, each term is non-variational, so the equation is non-variational as it stands.

If we want to impose the conditions on the variational vector fields, it provides that it has constant components. Such variations do not impose any constraints on solutions on $u$. Therefore, the KdV equation is not variational (as it stands) in the general sense. One can however make it variational by a simple substitution $u=\partial_{x}\phi$ that increases the number of derivatives in each term to an even number, see \cite{FunctionalExteriorCalculus-Book}. 
\end{Example}

Finally, we find properties of the variationality obstruction evaluated at the solution of the full problem. To start with, using vertical vector field $X$ on the jet space, we can extend the Cartan formula to define vertical Lie derivative $\hat{\mathcal{L}}_{X}$ by the commutator
\begin{equation}
 \hat{\mathcal{L}}_{X}\hat{\alpha} = [X\lrcorner, \hat{d}]\hat{\alpha},
 \label{Eq_CartanFormulaVertical}
\end{equation}
for some vertical form/functional form $\hat{\alpha}$. Applying this to an Euler functional one-form $\Omega$ that is variational ($\hat{d}\Omega = 0$), we have,
\begin{equation}
 \hat{\mathcal{L}}_{X}\Omega = \hat{d} (X\lrcorner \Omega). 
 \label{Eq_LvsOmegaVariational}
\end{equation}
We denote by $\Phi$ the solution manifold of $\Omega=0$, i.e., $\Phi^{*}\Omega=0$. By pulling-back along $\Phi$ we vanish both sides of (\ref{Eq_LvsOmegaVariational}). It can be seen by using $\Phi^{*}\hat{\mathcal{L}}_{X} = \hat{\mathcal{L}}_{\Phi^{*}X}\Phi^{*}$, where $\Phi^{*}X = d\Phi^{-1}(X\circ \Phi)$ is the pull-back of a vector field. We, therefore, arrived at
\begin{Proposition}
 For an Euler functional form $\Omega$ the equation (\ref{Eq_LvsOmegaVariational}) is fulfilled. This becomes a trivial constraint when pulled back on the solution manifold $\Omega=0$. 
\end{Proposition}

For non-variational case of $\Omega$, i.e., for $\hat{d}\Omega = \gamma \neq 0$ the equation (\ref{Eq_LvsOmegaVariational}) when pulled-back to the Euler-Lagrange solution manifold by $\Phi$ gives
\begin{equation}
 (\Phi^{*}X) \lrcorner \Phi^{*}\gamma = 0.
\end{equation}
The constraint must be fulfilled by the variation $X$ at the solution manifold given by $\Phi$.

\section{Conclusions}

Introducing the theory of antiexact forms into the framework of functional forms allows one to interpret existing results in a new way and provide new ones. We connect the (local) theory of functional exterior forms of the calculus of variations with the antiexact forms. This allows for defining the variational problem for some equations that are not variational and do not have a variational multiplier. The proposed optimization problem opens new perspectives for research.


\appendix

\section*{Acknowledgments}

This research was supported by the GACR grant GA22-00091S, the grant 8J20DE004 of the Ministry of Education, Youth and Sports of the CR, and Masaryk University grant MUNI/A/1092/2021. RK also thank the SyMat COST Action (CA18223) for partial support.

I want to thank prof. Vladimir Matveev and all of his group in Jena for the invitation and inspiring discussions. I also want to thank Josef \v{S}ilhan for continuous support.




\end{document}